\documentclass{juliacon}
\newcommand{\ts}{\textsuperscript}
\hypersetup{hidelinks}

\begin{document}


\title{Distributed Parallelization of xPU Stencil Computations in Julia}

\author[1]{Samuel Omlin}
\author[2, 3]{Ludovic R\"ass}
\author[2, 3]{Ivan Utkin}
\affil[1]{Swiss National Supercomputing Centre (CSCS), ETH Zurich, Lugano, Switzerland}
\affil[2]{Laboratory of Hydraulics, Hydrology and Glaciology (VAW), ETH Zurich, Zurich, Switzerland}
\affil[3]{Swiss Federal Institute for Forest, Snow and Landscape Research (WSL), Birmensdorf, Switzerland}


\hypersetup{
pdftitle = {Distributed Parallelization of xPU Stencil Computations in Julia},
pdfsubject = {JuliaCon 2019 Proceedings},
pdfauthor = {Samuel Omlin, Ludovic R\"ass, Ivan Utkin},
pdfkeywords = {Julia, Distributed Parallelization, xPU, GPU, Supercomputing, Stencil Computations, Staggered Grid},
}

\maketitle

\begin{abstract}

We present a straightforward approach for distributed parallelization of stencil-based xPU applications on a regular staggered grid, which is instantiated in the package \texttt{ImplicitGlobalGrid.jl}. The approach allows to leverage remote direct memory access and enables close to ideal weak scaling of real-world applications on thousands of GPUs. The communication costs can be easily hidden behind computation.

\end{abstract}

\section{Introduction}
 In light of the high pace of hardware evolution since the dawn of the 21\ts{st} century, the HPC community has identified the 3 ``P''s - (scalable) Performance, (performance) Portability and Productivity - as fundamental requirements for today's and tomorrow's software development. The approach and package development presented in this paper responds to each of the 3 ``P''s. We present an approach for automatic and architecture-agnostic distributed parallelization of stencil-based xPU applications on a regular staggered grid (with xPU we refer simultaneously to GPU and CPU in this paper). 

\section{Approach}
The here presented approach renders the distributed parallelization of stencil-based xPU applications on a regular staggered grid almost trivial. We have instantiated the approach in the Julia package \texttt{ImplicitGlobalGrid.jl}. A highlight in the design of ImplicitGlobalGrid is the automatic implicit creation of the global computational grid based on the number of processes the application is run with (and based on the process topology, which can be explicitly chosen by the user or automatically defined). As a consequence, the user only needs to write a code to solve his problem on one xPU (local grid); then, as little as three functions can be enough to transform a single xPU application into a massively scaling multi-xPU application: a first function creates the implicit global staggered grid, a second function performs a halo update on it, and a third function finalizes the global grid. Fig.~\ref{fig:code} shows a stencil-based 3-D heat diffusion xPU solver, where distributed parallelization is achieved with these three ImplicitGlobalGrid functions (lines 23, 38 and 43) plus some additional functions to query the size of the global grid (lines 24-26; note that shared memory parallelization is performed with ParallelStencil \cite{parallelstencil2022}).

\begin{figure}[t]
\begin{lstlisting}[language = Julia, numbers=left, numberstyle=\tiny\color{gray}]
using ImplicitGlobalGrid
using ParallelStencil
using ParallelStencil.FiniteDifferences3D
@init_parallel_stencil(CUDA, Float64, 3)

@parallel function step!(T2,T,Ci,lam,dt,dx,dy,dz)
    @inn(T2) = @inn(T) + dt*(
        lam*@inn(Ci)*(@d2_xi(T)/dx^2 + 
                      @d2_yi(T)/dy^2 + 
                      @d2_zi(T)/dz^2 ) )
    return
end

function diffusion3D()
    # Physics
    lam      = 1.0           #Thermal conductivity
    c0       = 2.0           #Heat capacity
    lx=ly=lz = 1.0           #Domain length x|y|z

    # Numerics
    nx=ny=nz = 512           #Nb gridpoints x|y|z
    nt       = 100           #Nb time steps
    me,      = init_global_grid(nx, ny, nz)
    dx       = lx/(nx_g()-1) #Space step in x
    dy       = ly/(ny_g()-1) #Space step in y
    dz       = lz/(nz_g()-1) #Space step in z

    # Initial conditions
    T  = @ones(nx,ny,nz).*1.7 #Temperature
    T2 = copy(T)              #Temperature (2nd)
    Ci = @ones(nx,ny,nz)./c0  #1/Heat capacity

    # Time loop
    dt = min(dx^2,dy^2,dz^2)/lam/maximum(Ci)/6.1
    for it = 1:nt
        @hide_communication (16, 2, 2) begin
            @parallel step!(T2,T,Ci,lam,dt,dx,dy,dz)
            update_halo!(T2)
        end
        T, T2 = T2, T
    end

    finalize_global_grid()
end

diffusion3D()

\end{lstlisting}

    \caption{Stencil-based 3-D heat diffusion xPU solver implemented using ImplicitGlobalGrid and ParallelStencil.}
	\label{fig:code}
\end{figure}

ImplicitGlobalGrid relies on \texttt{MPI.jl} \cite{byrne2021mpi} to perform halo updates close to hardware limits. For GPU applications, ImplicitGlobalGrid leverages remote direct memory access when CUDA- or ROCm-aware MPI is available and, otherwise, uses highly optimized asynchronous data transfer routines to move the data through the hosts. In addition, pipelining is applied on all stages of the data transfers, improving the effective throughput between GPU and GPU. Low level management of memory, CUDA streams, ROCm queues and signals permits to efficiently reuse send and receive buffers and streams or queues and signals throughout an application without putting the burden of their management to the user. Moreover, all data transfers are performed on non-blocking high-priority streams or queues, allowing to overlap the communication optimally with computation. \texttt{ParallelStencil.jl}, e.g., can do so with a simple macro call (Fig.~\ref{fig:code}, line 36).

ImplicitGlobalGrid is fully interoperable with \texttt{MPI.jl}. By default, it creates a Cartesian MPI communicator, which can be easily retrieved together with other MPI variables. Alternatively, an MPI communicator can be passed to ImplicitGlobalGrid for usage. As a result, ImplicitGlobalGrid's functionality can be seamlessly extended using \texttt{MPI.jl}.

The modular design of ImplicitGlobalGrid, which heavily relies on multiple dispatch, enables adding support for other hardware with little development effort. Support for AMD GPUs using the recently matured \texttt{AMDGPU.jl} package \cite{amdgpu_jl} has already been implemented as a result (Nvidia GPUs are supported using \texttt{CUDA.jl} \cite{besard2018effective}). ImplicitGlobalGrid supports at present distributed parallelization for CUDA- and ROCm-capable GPUs as well as for CPUs.

\section{Results}
\begin{figure}[t]
    \centerline{\includegraphics[width=8cm]{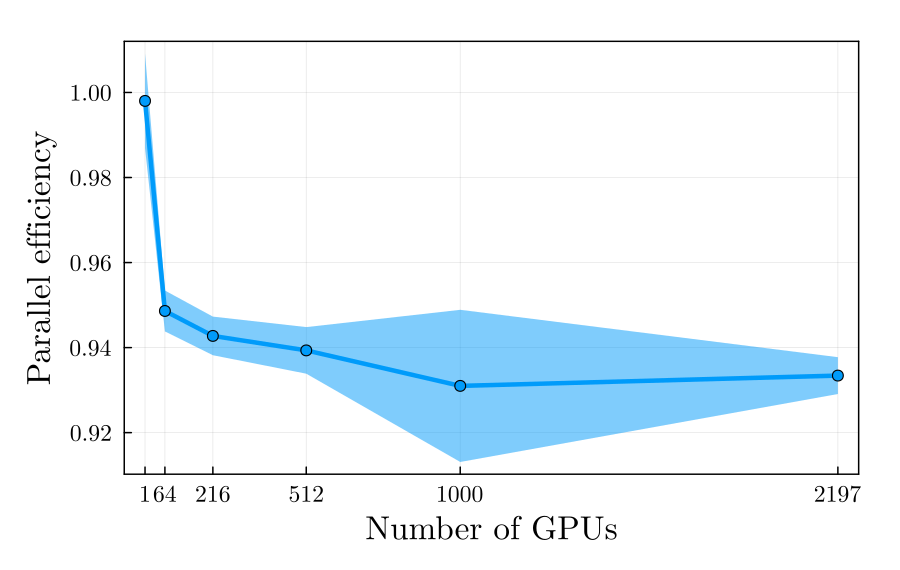}}
    \caption{Parallel weak scaling of the 3-D heat diffusion solver (Fig.~\ref{fig:code}) from $1$ to $2197$ Nvidia P100 GPUs on Piz Daint at CSCS. The blue surface visualizes the 95\% confidence interval of the reported medians (20 samples). The raw data and plotting script are available in \url{github.com/omlins/ImplicitGlobalGrid.jl/tree/master/paper}.}
	\label{fig:weak_scaling}
\end{figure}

\begin{figure}[t]
    \centerline{\includegraphics[width=8cm]{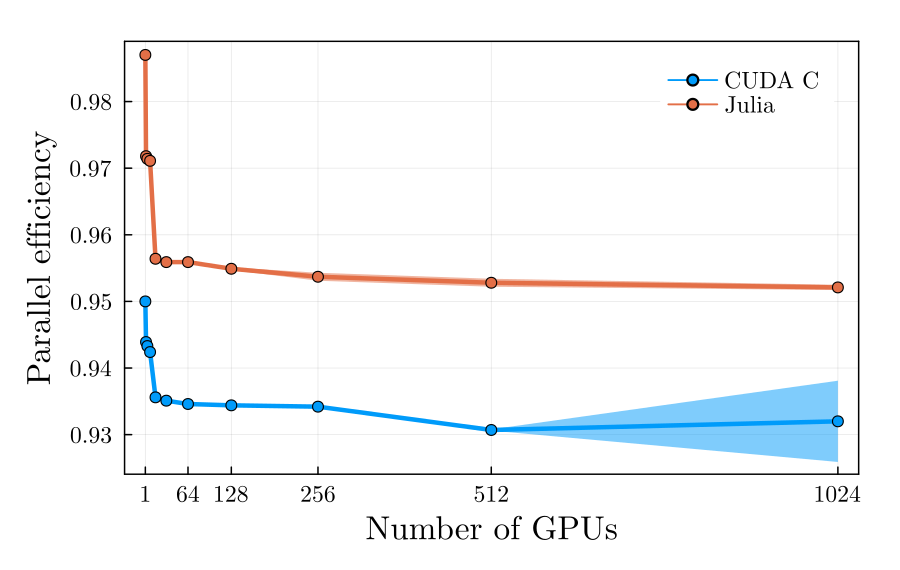}}
    \caption{Parallel weak scaling of the nonlinear 3-D poro-visco-elastic two-phase flow solver from $1$ to $1024$ Nvidia P100 GPUs on Piz Daint at CSCS (problem size per GPU is $382^\mathrm{3}$). The blue and orange surfaces visualize the 95\% confidence interval of the reported medians (20 samples). The raw data and plotting script are available in \url{github.com/omlins/ImplicitGlobalGrid.jl/tree/master/paper}.}
	\label{fig:weak_scaling_realworld}
\end{figure}

We here report the scaling achieved with the 3-D heat diffusion xPU solver (Fig.~\ref{fig:code}) on up to 2197 Nvidia Tesla P100 GPUs on the Piz Daint Supercomputer at the Swiss National Supercomputing Centre (Fig.~\ref{fig:weak_scaling}). We observe a parallel efficiency of 93\% on 2197 GPUs.
Moreover, we have employed ImplicitGlobalGrid and ParallelStencil for the parallelization of a solver for nonlinear 3-D poro-visco-elastic two-phase flow and have also conducted a weak scaling experiment on Piz Daint (Fig.~\ref{fig:weak_scaling_realworld}). We observe a parallel efficiency of over 95\% on up to 1024 GPUs. As a performance reference, the solver implemented in Julia achieved 90\% of the performance of the respective original solver written in CUDA C using MPI.

\section{Conclusions}
We have shown that ImplicitGlobalGrid enables scalable performance, performance portability and productivity and addresses the 3 ``P''s in all of its aspects. In addition, we have demonstrated the effectiveness and wide applicability of our approach within geosciences. Our approach is naturally in no sense limited to geosciences as distributed parallelization based on halo updates is employed in many scientific disciplines. We illustrated this in a recent contribution, where we showcased a quantum fluid dynamics solver using the nonlinear Gross-Pitaevski equation implemented with ImplicitGlobalGrid and ParallelStencil \cite{pasc21}.

\section{Acknowledgments}
We would like to thank Julian Samaroo (MIT) for his pro-active support for enabling AMDGPU in ImplicitGlobalGrid. This work was supported by a grant from the Swiss National Supercomputing Centre (CSCS) under project ID c23 through the Platform for Advanced Scientific Computing (PASC) program.


\bibliographystyle{juliacon}
\bibliography{ref.bib}

\end{document}